# Bone regenerative potential of mesenchymal stem cells on a micro-structured titanium processed by wire-type electric discharge machining

Yukimichi Tamaki*, Yu Kataoka, and Takashi Miyazaki

*Department of Oral Biomaterials and Technology, School of Dentistry, Showa University, 1-5-8 Hatanodai, Shinagawa-ku, Tokyo 142-8555, Japan*

**A new strategy with bone tissue engineering by mesenchymal stem cell transplantation on titanium implant has been dawn attention. The surface scaffold properties of titanium surface play an important role in bone regenerative potential of cells. The surface topography and chemistry are postulated to be two major factors increasing the scaffold properties of titanium implants. This study aimed to evaluate the osteogenic gene expression of mesenchymal stem cells on titanium processed by wire-type electric discharge machining. Some amount of roughness and distinctive irregular features were observed on titanium processed by wire-type electric discharge machining. The thickness of suboxide layer was concomitantly grown during the processing. Since the thickness of oxide film and micro-topography allowed an improvement of mRNA expression of cells, titanium processed by wire-type electric discharge machining is a promising candidate for mesenchymal stem cell based functional restoration of implants.**

\* contact to: tamaki@dent.showa-u.ac.jp

**Key words** Titanium, implant, surface, mesenchymal stem cell

## Introduction

The dental implant failure is frequently ascribed to large bone defect or surrounding poor bone quality. Functional restoration of bone following surgical interventions is a crucial step for the success and long term survival of orthopedic and dental implants[1, 2]. Although the bone graft is widely accepted as the standard in such surgical site, inherent donor-site limitations with respect to tissue rejection and disease transfer are of particular shortcomings in autografting[2, 3]. A new strategy with bone tissue engineering by mesenchymal stem cell (MSC) transplantation has been dawn attention[1, 4-7]. Titanium is a primary metallic biomaterial used in orthopedic and dental implants[8, 9]. The surface scaffold properties of titanium surface, namely osteogenic gene expression of adherent cells play an important role in the bone tissue engineering of implants[4, 7, 10, 11]. A large number of surface modification techniques have already existed, including commercial ones. Rough surfaces such as the sand-blasted, acid-etched titanium implant have generally proven superior to smooth surface in terms of promoting bone-cells and implant contact at the interface[12-14]. The authors have reported that the wire-type electric discharge machining (W-EDM) of titanium allowed a microstructured titanium surface with an irregular morphology. W-EDM processes electro-conductive materials by spark discharge generated between a narrow metal wire and the material through running pure distilled water. This computer associated technique enables extremely accurate complex sample shaping. It also yields an optimal micro textured surface during the processing[15].

Various thicknesses of the oxide film the type of crystallinity, and the amount of oxygen incorporated into the surfaces of differently processed titanium samples have been reported[13, 16-18]. The thickness of oxide film is another important determinant of osteoblast responses on titanium as the thickness of oxide film may improve the surface free energy that accumulates functional groups and osteogenic biomolecules at the surface. The surface topography and chemistry are postulated to be two major factors increasing the osteoconductivity of titanium implants. Titanium alters the thickness of the surface oxide film even in a simple heated setting. Since the titanium oxide film is sensitively altered by each processing, the authors hypothesized that W-EDM processes the thickness of oxide film and micro topography at the same time.

In this study, the chemical characteristics and surface topographies of titanium processed by W-EDM and titanium-coated W-EDM epoxy resin replicas were evaluated. The scaffold property of the samples was evaluated by the mRNA expression of MSC onto the samples.

## Materials and Methods

*Specimen Preparation*
JIS grade 2 titanium (KS-50, Kobe Steel, Kobe, Japan) was used as the starting material. The smooth surfaces of titanium plates, 10 x 10 x 1.0 mm, were gradually ground with waterproof polishing papers from #500 to #1200 under running water. They were then polished with alumina particles until an average diameter of 0.3 μm. W-EDM samples (W-EDM-Ti) of the same dimensions were processed under conditions of 6.5 μsecond τ off (pulse off time) and 0.65 μsecond τ on (pulse on time) at 15 amps $I_P$ (peak current) and 90 V (non-load voltage) (LS 350X, Japan). The samples were ultrasonically cleaned in acetone, detergent solution (7X, ICN), and pure distilled water for 15 min of each cleaning process. The specimens were then dried and stored in a desiccator for 24 hrs in 50% humidity and at a temperature of 23°C.

*Preparation of replica surface topography*
The method for titanium coated replicas was given in the paper written by Burunett DM et al[19, 20]. Impressions of polished titanium and W-EDM surfaces were made with vinyl polysiloxane impression material (PROVIL® *novo Light*, Heraeus Kulzer, Dormagen, Germany). Vinyl polysiloxane negative replicas were used to cast epoxy-resin (EPO-TEK 302-3, Epoxy Technology, Bellerica, MA) positive replicas of the surfaces. The epoxy replicas of polished titanium (Ti) and W-EDM were cleaned with in a detergent (7X) (ICN Biomedicals, Inc., Costa Mesa, CA), baked at 60°C for 4 days, and sputter-coated (IB-2, Eiko Engineering, Tokyo, Japan) with 50 nm of Ti.

*Micro topography*
The micro topography of the samples was obtained using a scanning electron microscope (S-2360N, HITACHI, Hitachi, Japan). The average roughness of the samples was evaluated by a surface texture measuring instrument (SURFCOM, 480A, TOKYO SEIMITUS, Tokyo, Japan). The results were expressed as the mean ± SD of 6 specimens (n = 6), and analyzed statistically by analysis of variance (ANOVA).

*Surface characterization*
XRD
The crystal phases of the samples were detected by TF-XRD (XRD-6100, SHIMADZU, Kyoto, Japan) with CuKα radiation. XRD was operated at 40 kV, 40 mA with a scanning speed of 0.02°/4 sec and a scanning range of 20-50°.

XPS
The surfaces of each specimen were characterized by X-ray photoelectron spectroscopy (XPS) (ESCA-3400, SHIMADZU, Kyoto, Japan). High-resolution spectra from a wide scan were analyzed using MgKα radiation. A 20-mA emission current and 8-kV accelerated voltage were applied for this analysis. The binding energies for each spectrum were calibrated with the C1s spectrum of 285.0 eV. The results were analyzed statistically by Student's t tests. Significant differences were considered to exist when $p < 0.01$ (n=6).

*Expression of mRNA on the samples*
Rat MSCs (RIKEN Cell Bank, Tsukuba, Japan) were cultured on the titanium samples with α-MEM (Sigma, Tokyo, Japan) containing 10 % FBS, dexamethasone, ascorbic acid 2 phosphate and gryceropohosphate at 37 °C and 5 % $CO_2$ for 1 week. Expressions of integrin αV, osteopontin, and Cbfa-1 were obtained by RT-PCR

**Results**
*Surface topographies*
The surface topography of Ti appeared to be smoother than those of W-EDM-Ti and W-EDM replicas (Fig.1). Some amount of roughness and distinctive irregular features were observed on W-EDM-Ti and W-EDM replicas. Average surface roughnesses, Ra, of the samples were the same; the average for W-EDM-Ti was 19.5±0.5μm, and for W-EDM replicas it was 19.6±0.4μm (Fig.1). No significant differences ($p > 0.01$) were detected between W-EDM-Ti and W-EDM replicas.

*Surface characterization*
X-ray diffraction patterns of specimen are shown in Fig.2. Typical XRD patterns of titanium were detected similarly on Ti and W-EDM replicas. A peak of 42.5 2θ (Cakα/degree) attributable to TiO appeared to be increased on W-EDM-Ti.
High resolution spectra of Ti2p on each sample are shown. The binding energies of each spectrum between Ti and W-EDM replicas showed no significant differences ($p<0.01$). The shoulder peak of Ti2p at 457.0 eV attributable to TiO was increased on W-EDM-Ti compared to Ti and W-EDM replicas (not shown).

*Expression of mRNA*
Integrin αV expressions were detected only on W-EDM-Ti even after 6 hrs cultivation. The mRNA expression of osteopontin was observed after 1 wk on all samples. The expression of cbfa-1 on W-EDM-Ti was observed after 1 wk cultivation while that of W-EDM replica was observed after 2 wks. The expression of cbfa-1 on Ti was observed after 3 wks culture.

**Discussion**
The present investigation of surface characterization revealed that the superficial chemistry of Ti and W-EDM replicas was consistent. Since the surface topographies of W-EDM-Ti and W-EDM replicas showed no significant differences, W-EDM replicas could solely replicate the micro topography effect of W-EDM-Ti on *in vitro* cellular responses. Both surface characterizations demonstrated that an increasing TiO peak was detected only on W-EDM-Ti samples. Oxygen incorporated into the W-EDM-Ti surface increased during processing. Thus, W-EDM processing could alter the superficial chemistry as well as the micro topography of titanium surfaces.

Cells interact with titanium substrate via integrin binding to extra cellular matrix proteins[21, 22]. The first phase integrin mediated cell adhesion mechanism is essential to achieve the successful anchorage of endosseous implants[23]. The initial adhesion of cells is mediated by specific membrane receptors, mainly integrins αV[21, 22]. These recognize the biding domains of the RGD sequence of adsorbed serum proteins such as fibronectin or vitronectin on titanium surfaces. A thickness of TiO suboxide layer increases the surface free energy that accumulates functional groups and osteogenic biomolecules when exposed in biological fluids. Since the expression of integrin αV were detected only on W-EDM-Ti, even after 6 hrs cultivation, the increased TiO layer formed on the W-EDM-Ti increased the amount of cell binding serum proteins that could be affected by the increased surface free energy.

The Cbfa-1 is a related gene which indicates differentiation from MSCs to osteoblasts[24]. Recent studies have indicated that the different surface topographies of titanium distinctly modulate cell morphology of fibroblasts and osteoblastic cells at surfaces[10, 25]. Morphological features of osteoblastic cells markedly activated cytoplasmic stress fiber formation that is associated with a more differentiated cellular phenotype. The present study showed some amount of roughness and distinctive irregular features on W-EDM-Ti and W-EDM replicas. Thus, Cbfa-1 expression on W-EDM-Ti and W-EDM replicas was up-regulated compared to that on smooth Ti samples.

Both surface chemistry and topography of W-EDM-Ti up-regulated the mRNA expression of MSCs. W-EDM processing on titanium implants therefore enhances scaffold properties of MSCs to the surface. In conclusion, the thickness of oxide film and micro-topography performed on W-EDM-Ti has scaffold properties toward the MSCs. W-EDM is a promising candidate for MSC based functional restoration surrounding implants.

**ACKNOWLEDGEMENTS**
This work was supported by MEXT, Haiteku (2007), a Grant-in-Aid for Scientific Research (B) from the Japan Society for the Promotion of Science, and a Grant-in-Aid for the Encouragement of Young Scientists (B) from The Ministry of Education, Culture, Sports, Science and Technology of Japan.

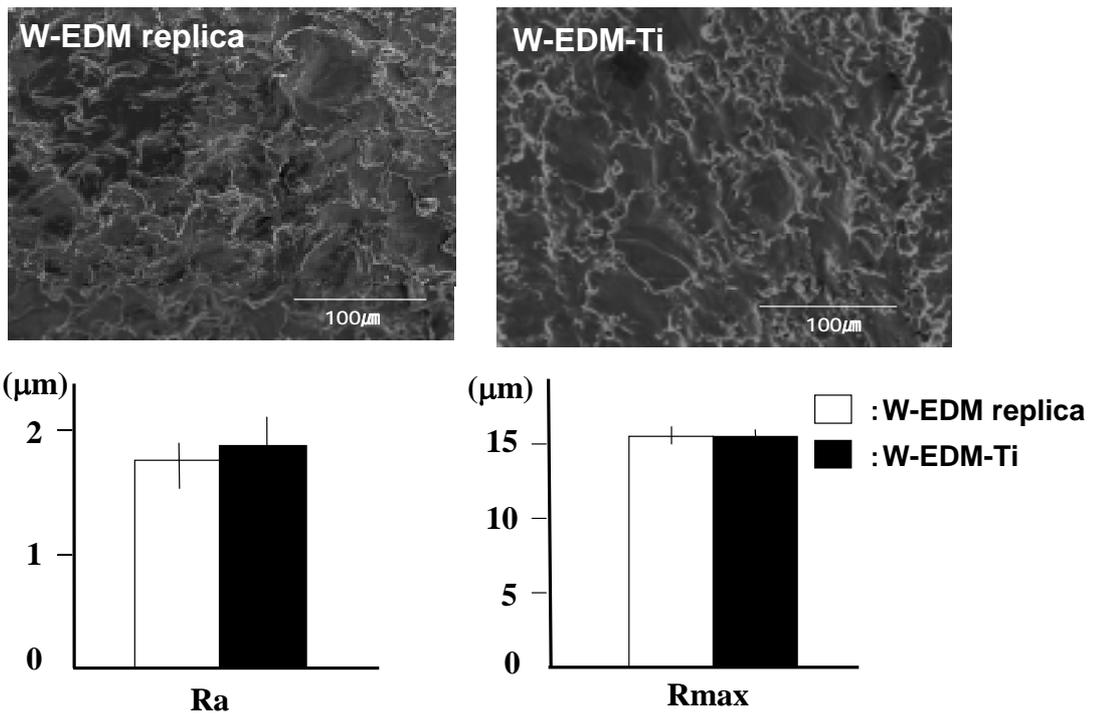

Figure 1 Surface topography of W-EDM-Ti and W-EDM replica

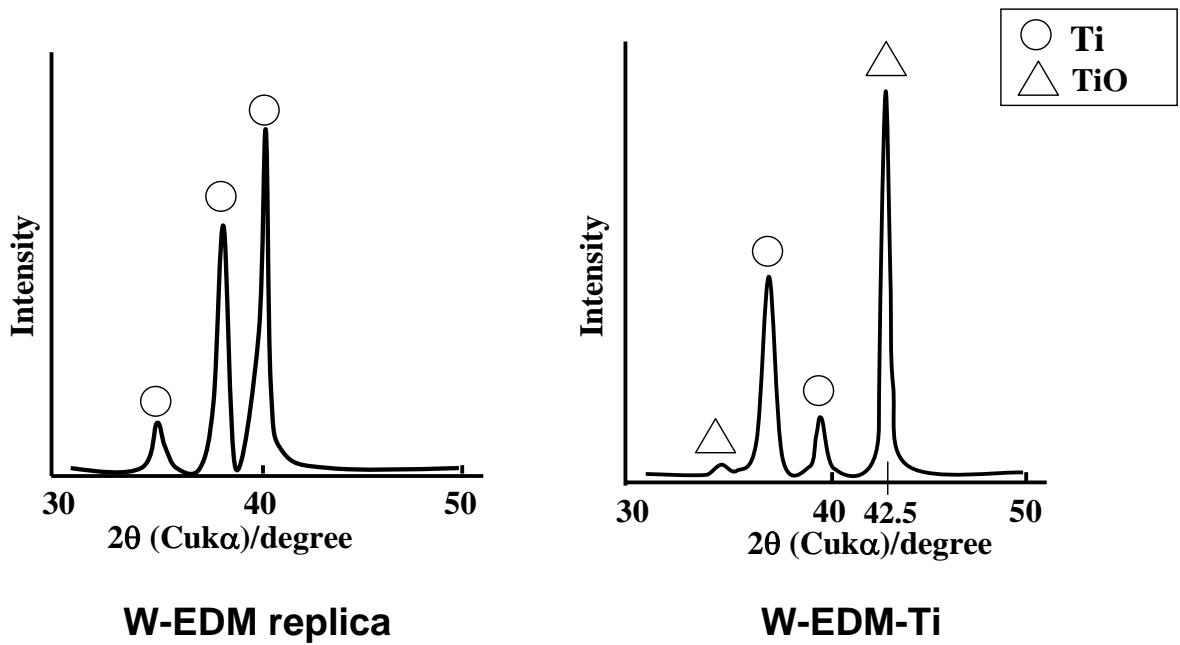

Figure 2 XRD spectra of W-EDM-Ti and W-EDM replica

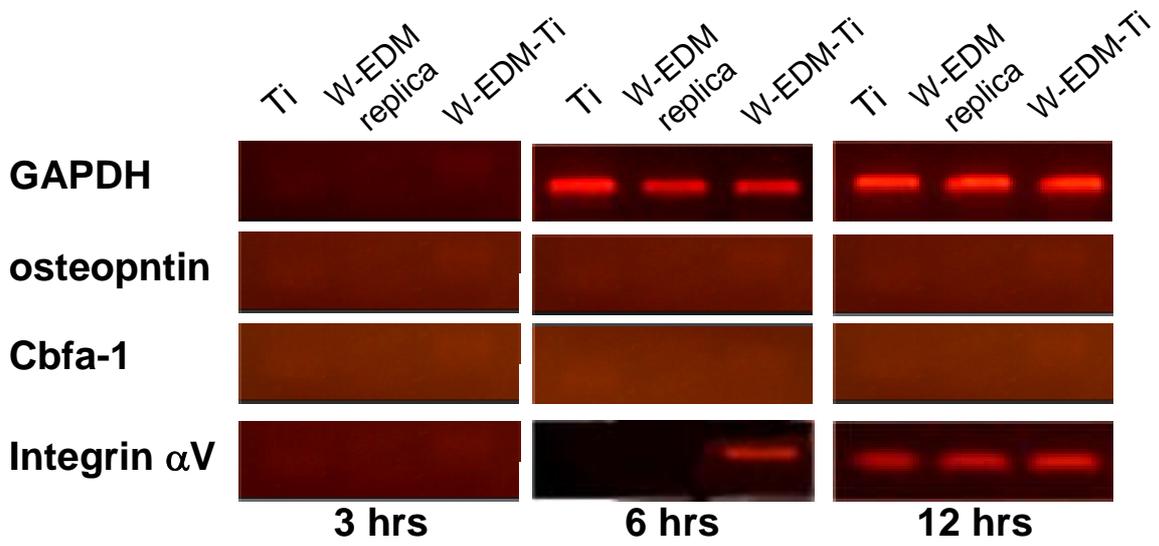

Figure 3  mRNA expression of MSCs on the samples for 12 hrs.

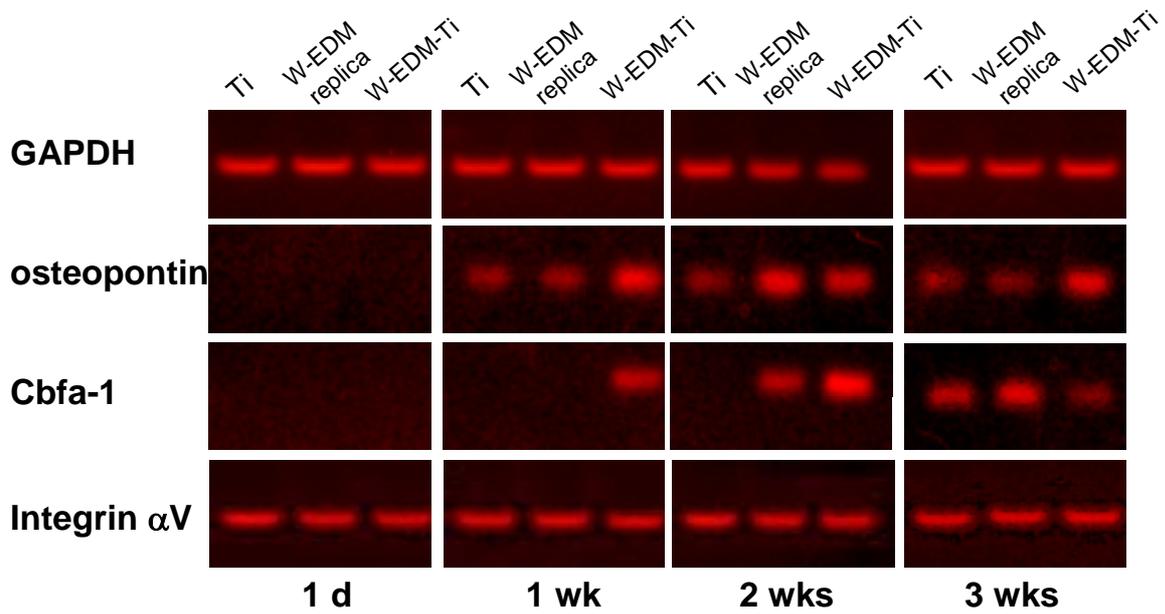

Figure 4 mRNA expression of MSCs on the samples for 3 wks